# Elastic anisotropy of lizardite at subduction zone conditions

**Xin Deng[1], Chenxing Luo[2], Renata M. Wentzcovitch[2,3,4], Geoffrey A. Abers[5], and Zhongqing Wu[1,6,7*]**

[1] Laboratory of Seismology and Physics of Earth's Interior, School of Earth and Space Sciences, University of Science and Technology of China, Hefei, Anhui 230026, China.

[2] Department of Applied Physics and Applied Mathematics, Columbia University, New York, NY 10027, USA.

[3] Department of Earth and Environmental Sciences, Columbia University, New York, NY 10027, USA.

[4] Lamont–Doherty Earth Observatory, Columbia University, Palisades, NY 10964, USA.

[5] Department of Earth and Atmospheric Sciences, Cornell University, Ithaca, NY, 14850, USA.

[6] CAS Center for Excellence in Comparative Planetology, China, Anhui 233500, China.

[7] National Geophysical Observatory at Mengcheng, University of Science and Technology of China, Hefei, China.

Corresponding author: Zhongqing Wu (wuzq10@ustc.edu.cn)

**Key Points:**

- The elastic properties of lizardite are determined at *P-T* conditions corresponding to subduction zones using *ab initio* calculations

- The elastic anisotropy of lizardite for S wave is almost double that of antigorite

- The large elastic anisotropy of lizardite can account for the observed shear-wave splitting in the subducting slabs near trenches

## Abstract

Subduction zones transport water into Earth's deep interior through slab subduction. Serpentine minerals, the primary hydration product of ultramafic peridotite, are abundant in most subduction zones. Characterization of their high-temperature elasticity, particularly their anisotropy, will help us better estimate the extent of mantle serpentinization and the Earth's deep water cycle. Lizardite, the low-temperature polymorph of serpentine, is stable under the *P-T* conditions of cold subduction slabs (< 260°C at 2 GPa), and its high-temperature elasticity remains unknown. Here we report *ab initio* elasticity and acoustic wave velocities of lizardite at *P-T* conditions of subduction zones. Our static results agree with previous studies. Its high-temperature velocities are much higher than previous experimental-based lizardite estimates with chrysotile but closer to antigorite velocities. The elastic anisotropy of lizardite is much larger than that of antigorite and could better account for the observed large shear-wave splitting in some cold slabs such as Tonga.

## Plain Language Summary

Serpentine minerals are crucial water carriers in subduction zones. Their layered structure is responsible for their strong elastic anisotropy. However, their elastic properties, especially the anisotropy at high-temperature conditions remain unclear. In this study, we determined the elasticity of lizardite, the low-temperature polymorph of serpentine, at *P-T* conditions of subduction zones using *ab initio* calculations. The resulting velocities agree with previous calculation studies but are much higher than previous experimentally-based estimates. Lizardite has significant shear wave anisotropy. The large elastic anisotropy could account for the observed shear-wave splitting in the subducting slabs.

## 1 Introduction

Water plays a vital role in Earth's evolution and dynamics. It can be transported from the surface into the deep mantle through the subduction of slabs at plate boundaries (Faccenda et al., 2008; van Keken et al., 2011; Yang et al., 2017). It is also considered essential for the seismicity and magmatism of subduction zones (Cooper et al., 2020; Fujie et al., 2018) since fluids can reduce the melting point of rocks (Sobolev et al., 2019) and the effective normal stress of faults (Mibe et al., 2011; Ranero et al., 2003). Before subduction, bending-related faulting of slabs in trenches' outer rise region is a critical process for slab hydration (Grevemeyer et al., 2018; Ranero et al., 2003; Wang et al., 2022). Fluids enter the crust and mantle beneath it easily through the pathways provided by bend faults (Grevemeyer et al., 2005; Ivandic et al., 2008; Ranero et al., 2005) and then react with the mantle peridotite to hydrate slabs (Fujie et al., 2013; Grevemeyer et al., 2007; Ivandic et al., 2008).

The addition of water into the mantle composed of olivine and pyroxene may generate a variety of hydrous minerals such as serpentine (antigorite, lizardite, chrysotile), brucite, talc, and other secondary minerals like magnetite, depending on temperature, pressure, and bulk composition (Evans, 1977; Evans et al., 2013; Hyndman & Peacock, 2003; Janecky & Seyfried, 1986; Manning, 1995). The most abundant hydrous minerals in the altered mantle are serpentine minerals ($Mg_3Si_2O_5(OH)_4$), which contain ~13 wt% water. Mantle serpentinization alters the composition of the mineral assemblage and has significant effects on the physical and mechanical properties of the mantle, including a decrease in seismic velocities, an increase in seismic anisotropy, an increase in the Poisson's ratio, and a reduction in density (Hyndman & Peacock, 2003). Seismic array studies have revealed low velocity and high anisotropy structures in subduction zones (Arnulf et al., 2022; Cai et al., 2018; Long & van der Hilst, 2006; Long & Silver, 2008; Nakajima & Hasegawa, 2004; Wang et al., 2022; Wiens et al., 2008), suggesting the presence of serpentine minerals in subducting slabs (Arnulf et al., 2022; Cai et al., 2018; Cooper et al., 2020; Hyndman & Peacock, 2003; Wang et al., 2022). Since the elasticity of serpentine minerals might reveal the extent of serpentinization, they are essential for understanding the Earth's deep water cycle.

Lizardite, antigorite, and chrysotile are the three serpentine polymorphs with an approximate composition of $Mg_3Si_2O_5(OH)_4$. Chrysotile is a metastable phase relative to lizardite and antigorite. Besides, it is favored in isotropic stress microenvironments (Andreani et al., 2008; Evans, 2004). Lizardite is the most stable polymorph under ambient conditions and transforms to antigorite at high pressures and temperatures (Evans, 2004; Evans et al., 2013; Ghaderi et al., 2015; Reynard, 2013; Schwartz et al., 2013). As shown in Fig. 1a, the transition temperature at 0 GPa is 340 °C and drops to ~260 °C at 2 GPa, which suggests that lizardite is stable at many cold subduction zones in the West Pacific, e.g., Tonga, Honshu, and Mariana, at a depth of 60 km (Ghaderi et al., 2015; Syracuse et al., 2010). Therefore, the high-temperature elasticity of lizardite, especially the anisotropy, is of great significance for understanding seismic anisotropy in subduction zones, the volume of serpentinized mantle and related water content, and ultimately the Earth's deep water cycle. The equation of state of lizardite has been measured using X-ray diffraction at high pressure and room temperature (Hilairet et al., 2006a). However, the elasticity of lizardite has only been investigated by atomistic simulations (Auzende et al., 2006) and first-principles calculations at static conditions (Ghaderi et al., 2015; Mookherjee & Stixrude, 2009; Reynard et al., 2007; Tsuchiya, 2013). Measurements on these phases are challenging because of difficulties in preparation of phyllosilicates for experimental techniques such as Brillouin spectroscopy and the demanding nature of high temperature *ab initio* elasticity calculations. Despite the importance of lizardite for interpreting subduction zones' velocity structures and dynamics, little is known about its elastic properties at high pressures and temperatures.

We obtain the elastic properties of lizardite at high pressures and temperatures using *ab initio* calculations based on density functional theory (DFT) to address these fundamental questions. The resulting velocities are higher than those obtained in previous experiments but comparable to those in previous calculations (Auzende et al., 2006; Christensen, 1966; 2004; Ghaderi et al., 2015; Hilairet et al., 2006a; Ji et al., 2013; Mookherjee & Stixrude, 2009; Reynard et al., 2007; Tsuchiya, 2013). The elastic anisotropy of lizardite is large and sufficient to explain the seismic delay time observed in subduction zones such as Tonga and Middle America (Faccenda et al., 2008; Katayama et al., 2009; Long & van der Hilst, 2006; Long & Silver, 2008).

## 2 Methods

The elastic coefficient tensor under high pressure and temperature can be written as a relation with the Helmholtz free energy under isothermal conditions (Barron & Klein, 1965):

$$c_{ijkl}^T = \frac{1}{V}\left(\frac{\partial^2 F}{\partial e_{ij}\partial e_{kl}}\right) + \frac{1}{2}P\big(2\delta_{ij}\delta_{kl} - \delta_{il}\delta_{kj} - \delta_{ik}\delta_{jl}\big). \tag{1}$$

Here, $e_{ij}$ $(i, j = 1,3)$ represent the infinitesimal strains, $P$ represents the pressure, $\delta_{ij}$ represents the Kronecker delta symbol, and $F$ represents the Helmholtz free energy which can be calculated with quasiharmonic approximation (QHA):

$$F(e_{ij}, V, T) = U(e_{ij}, V) + \frac{1}{2}\sum_{q,m}\hbar\omega_{q,m}(e_{ij}, V) + k_B T \sum_{q,m}\ln\left\{1 - \exp\left[-\frac{\hbar\omega_{q,m}(e_{ij}, V)}{k_B T}\right]\right\}. \tag{2}$$

In Eq. (2), $\omega_{q,m}$ represents the vibrational frequency of the $m$-th normal mode with the phonon wave vector $q$. $V$ and $T$ represent the equilibrium volume and temperature. $\hbar$ and $k_B$ are the reduced Planck and Boltzmann constants, respectively. The three terms on the right side of Eq. (2) correspond to the static internal, zero-point, and vibrational energy contributions. To get the full elastic constant tensor, the vibrational density of states (VDoSs) and free energy of several strained configurations are needed in the conventional methods. Instead, the method used in this study (Luo et al., 2021; Wu & Wentzcovitch, 2011; Zou et al., 2018) only requires knowledge of the volume dependence of vibrational properties under hydrostatic pressure, avoiding the VDoSs calculations for the configurations under strains. This method involves a moderate computational workload, no more than 10% of those of conventional ones. It has been successfully adopted to obtain thermoelastic tensors of numerous minerals (Duan et al., 2019; Hao et al., 2019; Marcondes et al., 2015; Núñez-Valdez et al., 2011; Núñez-Valdez et al., 2013; Núñez Valdez et al., 2012a; Núñez Valdez et al., 2012b; Shukla et al., 2016; Shukla et al., 2019; Shukla et al., 2015; Wang et al., 2020; Wang et al., 2021; Wang et al., 2019; Wu et al., 2013). Then the adiabatic bulk modulus $K_S$ and shear modulus $G$ were calculated using the Voigt-Reuss-Hill averages of the elastic tensor. Compressional and shear velocities were derived from the equations:

$$V_P = \sqrt{\left(K_S + \frac{4}{3}G\right)\big/\rho}, V_S = \sqrt{G/\rho}. \tag{3}$$

Calculations in this study were performed with the Quantum ESPRESSO package (Giannozzi et al., 2009) based on the DFT and the **cij** package (Luo et al., 2021). Both the local density approximation (LDA) (Kohn & Sham, 1965; Perdew & Zunger, 1981) and generalized gradient approximation (GGA) (Perdew et al., 1996) were adopted as the exchange-correlation functional except density-functional-perturbation-theory (DFPT) calculations for VDoSs, where only LDA was used. Pseudopotentials for magnesium were generated using von Barth-Car's method (Karki et al., 2000; Umemoto et al., 2008). Troullier-Martins pseudopotentials (Troullier & Martins, 1991) were used for silicon and oxygen. And for hydrogen, an ultrasoft pseudopotential generated by the Vanderbilt method (Vanderbilt, 1990) was used. The plane wave kinetic energy cutoff was 70 Ry. The lizardite ($Mg_3Si_2O_5(OH)_4$) cell used in the study contains 18 atoms, and its structure was optimized with a $5 \times 5 \times 4$ $k$-point mesh using the variable cell-shape damped molecular dynamics approach (Wentzcovitch, 1991). The dynamical matrices for the optimized structures were calculated using DFPT (Baroni et al., 2001) on a $2 \times 2 \times 2$ $q$-point mesh then interpolated in a denser $10 \times 10 \times 8$ mesh to obtain the VDoSs. The static elastic constant tensor was obtained using stress-strain relations resulting from ±1.0% deformations for shear elastic coefficients ($C_{14}$, $C_{44}$, and $C_{66}$ in this study) or ±0.5% for others on the optimized unit cells followed by internal atomic relaxation.

## 3 Results

### 3.1 Heat capacity and Equation of State

The heat capacity at constant volume (isochoric) $C_V$ and at constant pressure (isobaric) $C_P$ within LDA are shown in Fig. S1. Calculated values agree well with experimental measurements (Robie & Hemingway, 1995), indicating the appropriateness of the QHA. QHA assumes that the phonon frequency is independent of temperature at fixed volume and ignores the intrinsic phonon-phonon interactions. The intrinsic anharmonicity, which increases with temperature and decreases with pressure, causes thermodynamic properties such as the heat capacity to deviate from the experimental results at high temperature (~1000 K for forsterite at 0 GPa). Elastic properties, such as bulk modulus, are in general, less affected by anharmonicity than the heat capacity (Wu & Wentzcovitch, 2009). The agreement between the calculation and measurement on the heat capacity suggests that the anharmonicity can be ignored for the elasticity of lizardite despite that some modes of lizardite show clear anharmonicity (Balan et al., 2007; Balan et al., 2021).

Fig. 2a shows the compression curves of lizardite. Our static and QHA results, up to 7 GPa and 1000 K are compared to previous X-ray diffraction measurements at room temperature (Hilairet et al., 2006a). The static volume at 0 GPa in this study is also compared to prior calculations and experimental measurements in Table 1 (Auzende et al., 2006; Ghaderi et al., 2015; Hilairet et al., 2006a; Mookherjee & Stixrude, 2009; Tsuchiya, 2013). The static volumes predicted within LDA are smaller than the room temperature experimental value, and those within GGA are larger than the experiment results. The static GGA results are closer to the measurements than the static LDA ones. The inclusion of vibrational contribution at 300 K expands the volume by ~1.5%. After its inclusion, LDA and GGA perform similarly in predicting the volume of lizardite at ambient conditions. The difference from measurements at 0 GPa and 300 K are 3.22% and 3.56%, respectively. In this study, LDA is adopted to compute the VDoSs for deriving the high-temperature properties of lizardite. The volume difference between our study and experimental measurements decreases with pressure, ranging from 3.22% at 0 GPa to 1.87% at 8 GPa (Fig. 2a), implying in a higher zero-pressure bulk modulus ($K_0$) of 78.9 GPa ($K_0' = 7.0$, $V_0 = 175.09$ Å$^3$) in this study compared with previous experiments ($K_0 = 71.0$ GPa, $K_0' = 3.2$, $V_0 = 180.92$ Å$^3$).

### 3.2 Elastic Properties

The elastic constants increase with pressure, while their pressure sensitivity (the magnitudes of $\partial C_{ij}/\partial P$) decreases with pressure. The static elastic constants ($C_{ij}$ in Voigt notation) of lizardite at 0 GPa in this study are listed in Table 1, among those from previous calculations (Auzende et al., 2006; Ghaderi et al., 2015; Mookherjee & Stixrude, 2009; Reynard et al., 2007; Tsuchiya, 2013). Our results overall agree well with previous studies. The calculated finite-temperature elastic coefficients of lizardite are shown in Fig. S2. Some elastic constants, like $C_{14}$, $C_{44}$, and $C_{66}$ are almost pressure-independent at high pressure, while most of the other elastic constants, especially $C_{11}$, $C_{12}$, and $C_{33}$ are strongly nonlinearly dependent on pressure. $C_{33}$ is only about a half of $C_{11}$ within LDA and even much smaller within GGA (Table 1). The lower $C_{33}$ indicates lizardite is more compressible along $c$ than along $a$ and $b$ and results from the weaker bonding along the [001] direction. This also reasonably explains why $C_{33}$ increases more rapidly than $C_{11}$ upon compression, as shown in previous calculations (Ghaderi et al., 2015; Mookherjee & Stixrude, 2009; Tsuchiya, 2013) and here.

The adiabatic bulk and shear moduli ($K_S$ and $G$) are calculated from elastic coefficients using the Voigt-Reuss-Hill averages (Hill, 1963) (Fig. 2b). Then the moduli and the density are fitted with the equation at different temperatures:

$$M = M_0 + \left(\frac{\partial M}{\partial P}\right)_0 \cdot P + \frac{1}{2}\left(\frac{\partial^2 M}{\partial P^2}\right)_0 \cdot P^2 + \frac{1}{6}\left(\frac{\partial^3 M}{\partial P^3}\right)_0 \cdot P^3 \qquad (4)$$

where $M$ refers to the bulk moduli ($K_S$), shear moduli ($G$), or the densities ($\rho$) up to 8 GPa. $M_0$ is the value of $M$ at 0 GPa. $\left(\frac{\partial M}{\partial P}\right)_0$, $\left(\frac{\partial^2 M}{\partial P^2}\right)_0$, and $\left(\frac{\partial^3 M}{\partial P^3}\right)_0$ are the first, second, and third derivatives of $M$ at 0 GPa (Table S1). The fitted data matches well with the original results (Fig. S3). The adiabatic bulk moduli increase with pressure but become less temperature-sensitive as pressure increases. The shear moduli are almost pressure-independent, despite a gentle decrease above 4 GPa.

### 3.3 Elastic Wave Velocity

Fig. 2c shows the pressure dependence of compressional velocity ($V_P$) and shear velocity ($V_S$) at different temperatures. Both velocities show negative temperature dependence, but the dependences are weaker at higher pressure. $V_P$ increases nonlinearly with pressure, while $V_S$ is insensitive to pressure. The static velocities of lizardite in this study are consistent with previous computational studies (Ghaderi et al., 2015; Mookherjee & Stixrude, 2009; Reynard et al., 2007). Because the differences between lizardite's and antigorite's static velocities are small (Ghaderi et al., 2015), velocity jumps caused by the lizardite → antigorite transition might be too subtle to be detectable. As shown in Fig. 2a, the volumes predicted with LDA are smaller than the experimental ones, while those predicted within GGA are larger than the experimental ones. Since LDA and GGA tend to overestimate and underestimate the bond strength, their estimation of the elasticities and velocities should bracket the experimental ones. Our static calculations indicate that the shear velocities predicted by GGA are ~7% smaller than LDA (3.89 km/s at ambient conditions), and the compressional wave velocities predicted by GGA are ~12% smaller than LDA result (7.12 km/s at ambient conditions). Both GGA and LDA predictions of the velocities are much higher than previous experimental estimations (roughly 2.3 km/s for $V_S$ and 4.9 km/s for $V_P$) (Christensen, 1966; 2004; Ji et al., 2013), and the inclusion of vibrational effects does not reconcile the difference. The lower velocity in the measurements might result from impurities, especially the presence of chrysotile detected in the experimental samples (Mookherjee & Stixrude, 2009; Reynard et al., 2007). While lizardite has a flat crystal structure, chrysotile forms multiwall nanotubes or nanoscrolls (Evans et al., 2013). Nanotube exterior and interior diameters are usually around 25 nm and 5 nm, respectively (Yada, 1971). Therefore, the porosity of the chrysotile nanotube packing has been suggested to be responsible for the relatively low velocity in the experiments (Mookherjee & Stixrude, 2009; Reynard et al., 2007). Besides, the particular texture of chrysotile nanotubes also allows damping of ultrasonic waves of a few hundred μm wavelength, resulting in the relatively low velocity in the ultrasonic measurements (Reynard et al., 2007). Chrysotile is favored by growth kinetics, but lizardite's greater thermodynamic stability suggests it should eventually overtake chrysotile and become the predominant serpentine form in cold subduction slab regions (Evans, 2004). Given that the seismic velocity of antigorite is only slightly higher than that of lizardite (Ghaderi et al., 2015), the shear velocity of serpentine minerals is around 3.81 km/s under shallow subduction zone conditions.

### 3.4 Anisotropy

The elastic wave velocities of a single crystal along different directions are determined using the Christoffel equation (Musgrave, 1970):

$$\left| C_{ijkl} n_j n_l - \rho V^2 \delta_{ik} \right| = 0 \tag{5}$$

where $C_{ijkl}$ is the elastic coefficient tensor, $n_j$ and $n_l$ represent the $j$th and $l$th component of the propagation direction $\mathbf{n} = (n_1, n_2, n_3)$, $\delta_{ik}$ is the Kronecker delta, $\rho$ and $V$ are the density and wave velocities, respectively. The velocities variations with the incidence angle between the seismic ray path and (001) plane of lizardite at 300 °C and 0.6 GPa are shown in Fig. 3a and 3d. The interlayer interactions are much more robust in LDA than in GGA. Therefore, LDA gives larger $C_{33}$ values and thus smaller directional variations in velocities compared to GGA. $V_P$ ranges from 5.87 km/s to 9.13 km/s within LDA (5.14 km/s to 9.38 km/s within GGA), with the velocity along [001] direction being much lower than that along [100] and [010] direction. $V_S$ shows no anisotropy along [001] direction. However, the difference between the two shear wave velocities ($V_{S1}$ and $V_{S2}$) in the same direction reaches its maximum along [100] and [010] axes.

Single crystal anisotropies indicating the maximum variation of wave velocities along different directions are defined by (Karki et al., 2001):

$$A_P = \frac{V_P^{\max} - V_P^{\min}}{V_P} \tag{6}$$

$$A_S = \frac{V_S^{\max} - V_S^{\min}}{V_S} \tag{7}$$

$$A_S^{\mathrm{po}} = \frac{V_{S1} - V_{S2}}{V_S} \tag{8}$$

where $A_P$ and $A_S$ are azimuthal anisotropies for compressional ($V_P$) and shear ($V_S$) waves, $A_S^{\mathrm{po}}$ is the polarization anisotropy representing the difference between the two shear wave velocities ($V_{S1}$ and $V_{S2}$) propagating in the same direction.

Lizardite has relatively strong elastic anisotropies (Fig. S4), especially for $V_S$. $A_S$ is 0.72 at ambient conditions and increases with pressure and temperature. $A_S$ is almost double that of antigorite (Mookherjee & Capitani, 2011). $A_S^{\mathrm{po}}$ behaves like $A_S$. $A_P$ is 0.44 at ambient conditions and decreases with pressure but increases with temperature. The $A_P$ is also larger than that of antigorite which decreases from 0.38 at 0 GPa to 0.30 at 2.5 GPa (Mookherjee & Capitani, 2011). The large anisotropies suggest that lizardite may play a significant role in the seismic anisotropies observed at subduction zones.

## 4 Geophysical Significance

Fluids react with peridotite in the slabs and produce serpentine at subduction zones. Among the three polymorphs of serpentine, chrysotile, lizardite, and antigorite, chrysotile is a metastable phase and not the most common one under shear stress conditions of subduction zones (Andreani et al., 2008; Evans, 2004). Therefore, serpentines should be predominantly in lizardite or antigorite form depending on the pressure and temperature conditions. For most cold subduction zones such as Honshu, Izu, Bonin, and Mariana, lizardite is the most stable phase up to at least 40 km depth and can even be stable to about 100 km depth at the coldest subduction zones like Tonga (Fig.1) (Ghaderi et al., 2015; Syracuse et al., 2010).

Deformation experiments reveal that antigorite prefers to orient parallel to the shear strain direction, which corresponds to elastic anisotropy characterized by a slow propagation direction [001] perpendicular to the shear plane (Escartín et al., 1997; Katayama et al., 2009; Norrell et al., 1989). Lizardite likely performs similarly under strain since it has very similar structure to antigorite (Evans et al., 2013). Therefore, seismic waves in lizardite would travel slower in the direction normal to the shear plane than in any other direction. Numerical models conjecture that bending-related faults in the slabs dip at about 30˚ from the vertical orientation at the trench and tend to dip toward the trench more than away (Ranero et al., 2003). These trenchward-dipping faults attain a near-vertical direction below forearc (Faccenda et al., 2008). Therefore, the shear deformation along the bend faults results in the trench-parallel near-vertical preferred orientation of lizardite in the slabs and produces strong seismic anisotropy distinguished by the fast directions parallel to the trenches.

Seismic anisotropy at subduction zones is intricate as it is composed of 4 layers: the sub-slab mantle, the slab, the mantle wedge, and the overlying plate (Long & Silver, 2008). The geometry and strength of seismic anisotropy in the sub-wedge regions can be determined by combining the local shear-wave splitting from sources near the slab surface and the shear-wave splitting of the vertically traveling SKS waves, generated by the conversion of compressional waves at the core-mantle boundary. In most subduction regions, the fast direction of sub-wedge splitting tends to be parallel to the trench (Long & Silver, 2008). The delay time in the sub-wedge regions shows noticeable spatial variations. For instance, the delay time is only 0.1-0.2 s in Ryukyu (Long & van der Hilst, 2005; 2006; Long & Silver, 2008), which can be easily explained by the crystal-preferred orientation of olivine in and below the slab. However, the delay time can reach 1-2 s in other subduction zones, such as Tonga and Middle America (Long & Silver, 2008), which is hard to be explained by olivine anisotropy because the required anisotropic layer (~100-200 km) full of deformed olivine must be unrealistically thick (Katayama et al., 2009; Mookherjee & Mainprice, 2014). Lizardite has significant shear wave anisotropy, even more prominent than that of antigorite (Fig. 3). A 10 km thick layer of lizardite could produce a large delay time of 2.2 s (3.0 s within GGA), compared with the 1.3 s delay time of 10 km antigorite layer (Mookherjee & Capitani, 2011), assuming perfect crystal alignment (Fig. 3f). Applying the typical ~40% alignment of serpentine fabrics from high-pressure deformation experiments (Katayama et al., 2009), a 10 km thick lizardite layer could result in a delay time of 0.8 s (1.0 s within GGA) (Fig. 1b). Therefore, our results put an upper bound on the anisotropy caused by serpentinization of the mantle.

In Tonga, where lizardite could be stable up to 110 km depth (Fig. 1a), the low-velocity layer in the subducting mantle is ~30 km thick (Contreras-Reyes et al., 2011). A low mantle velocity of 7.3 km/s (Contreras-Reyes et al., 2011) indicates a high degree of serpentinization over 60 vol% (Mookherjee & Capitani, 2011). Adopting 40% alignment of serpentine fabrics and 60 vol% serpentinization, the large delay time of ~1.8 s in the Tonga subduction zone (Long & Silver, 2008) requires an anisotropic layer of ~60 km if we use the elastic properties of antigorite (Mookherjee & Capitani, 2011) (Fig. 1b). Instead, if we use the elasticity of lizardite, only an anisotropic layer of ~30 km is required to explain the large delay time in Tonga, in agreement with the thickness of the low-velocity layer (Contreras-Reyes et al., 2011). The layer thickness could be even thinner if a higher degree of alignment of serpentine fabrics or higher degree of serpentinization is assumed. Therefore, the strong anisotropy of lizardite in the slabs with a pole of sheet plane oriented normally to the trench could account for the sizeable seismic delay time in subduction regions.

The serpentinized faults in the slabs are likely to be sampled by near-vertical teleseismic rays. Since lizardite is highly anisotropic and prefers a trench-parallel near-vertical orientation in faults with the [001] axis normal to the trench, the sampled $V_P$ of lizardite would be 9.13 km/s (9.38 km/s within GGA), assuming perfect crystal alignment (Fig. 3a,d), instead of the averaged velocity of 7.02 km/s (6.67 km/s within GGA) (Fig. 2c). In addition, the $V_P/V_S$ ratio of lizardite in this orientation ranges from 1.76–3.79 (1.77–4.63 within GGA) (Fig. 3b,e), wider than the isotropically averaged $V_P/V_S$ of 1.84 (1.84 within GGA). For more typical levels of alignment of serpentine fabrics (Katayama et al., 2009), the sampled $V_P$ would be 7.85 km/s (7.73 km/s within GGA) and the $V_P/V_S$ ratio in this orientation ranges from 1.81–2.40 (1.80–2.57 within GGA).

## 5 Conclusions

The elastic properties and acoustic velocities of lizardite are investigated at *P-T* conditions of subduction zones using *ab initio* calculations. Our static results agree well with previous studies. Our results show that the elastic coefficients $C_{14}$, $C_{44}$, and $C_{66}$ are almost pressure-independent at high pressure. Our calculations also show that $C_{11}$ is about twice as large as $C_{33}$, with the latter having a stronger pressure dependence. Our calculations corroborate previous computational studies that show the acoustic velocities of lizardite should be much larger than previous experiment-based velocity estimates of chrysotile-bearing lizardite and should be much closer to those of antigorite. The inclusion of vibrational effects does not reconcile the differences. Our results also suggest that lizardite has strong seismic anisotropy, even much stronger than antigorite.

Seismic studies reveal that shear-wave velocities in slabs are faster parallel to the trench in most subduction zones. It is challenging to explain the large delay time of 1–2 s in some subduction regions by invoking the anisotropy of olivine in the mantle wedge. However, this delay time could be reconciled by including highly anisotropic lizardite layers. A 10 km thick layer of pure lizardite could produce a large shear-wave delay time of 2.2–3.0 s, assuming perfect crystal alignment or 0.8–1.0 s, assuming a more typical level of alignment (~40%). With a 40% alignment of serpentine fabrics and 60 vol% serpentinization, a lizardite layer with the thickness of the low velocity layer (~30 km, Contreras-Reyes et al., 2011), instead of an antigorite layer with ~60 km thickness, can generate the large delay time of ~1.8 s in Tonga subduction zone (Long & Silver, 2008). Therefore, the strong anisotropy of lizardite could account for the observed significant delay time in the subducting slabs near trenches.

## Declaration of competing interest

The authors declare that they have no known competing financial interests or personal relationships that could have appeared to influence the work reported in this paper.

## Acknowledgments

This work is supported by the National Key R&D Program of China (2018YFA0702703), Natural Science Foundation of China (41925017), and the Fundamental Research Funds for the Central Universities (WK2080000144). RMW and CL have been supported by DOE grant DE-SC0019759. Computations were conducted in the Supercomputing Center of the University of Science and Technology of China and the Hefei Advanced Computing Center.

## Data Availability Statement

The authors comply with the AGU's data policy, and the datasets of this paper are available at zenodo: https://doi.org/10.5281/zenodo.4600382.

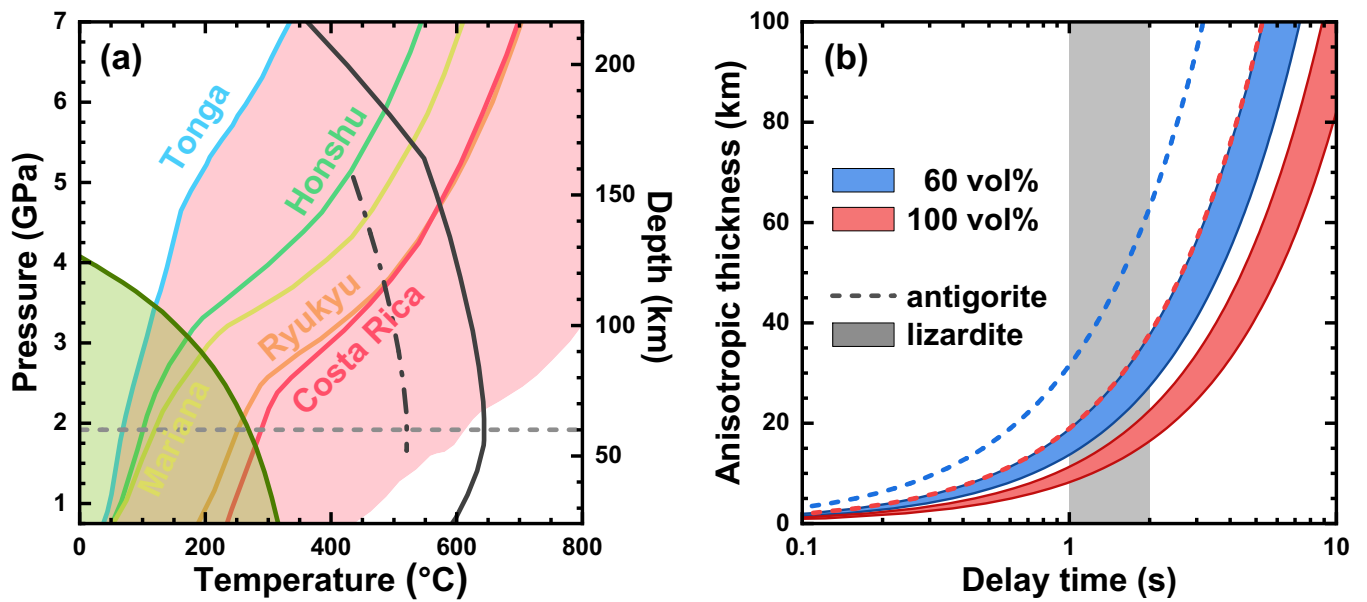

**Figure 1**. (a) Stability field of lizardite and *P-T* paths of the slab Moho. The dark green thick solid curve shows the phase boundary between lizardite and antigorite + brucite (Ghaderi et al., 2015) and the stability field of lizardite is shown as the green shaded area. The *P-T* paths of the slab Moho (7 km beneath the slab surface) for most subduction zones are shown as the pink area, and the *P-T* paths of the slab Moho for 5 reference subduction zones are presented with their locations labeled in the same color (Syracuse et al., 2010). The solid black curve represents the maximum stability of antigorite (Hilairet et al., 2006b; Nestola et al., 2009) and the dashed dotted black line shows the dehydration of antigorite under water absent conditions (Perrillat et al., 2005). Dashed gray line represents 60 km depth. (b) Relation between shear-wave delay time and required thickness of anisotropic layer, assuming lizardite (bands) and antigorite (dashed lines) with 40% alignment and different extent of serpentinization (blue: 60 vol% / red: 100 vol%) as sources. The gray shaded area indicates the large delay time of 1~2 s observed in several subdution zones.

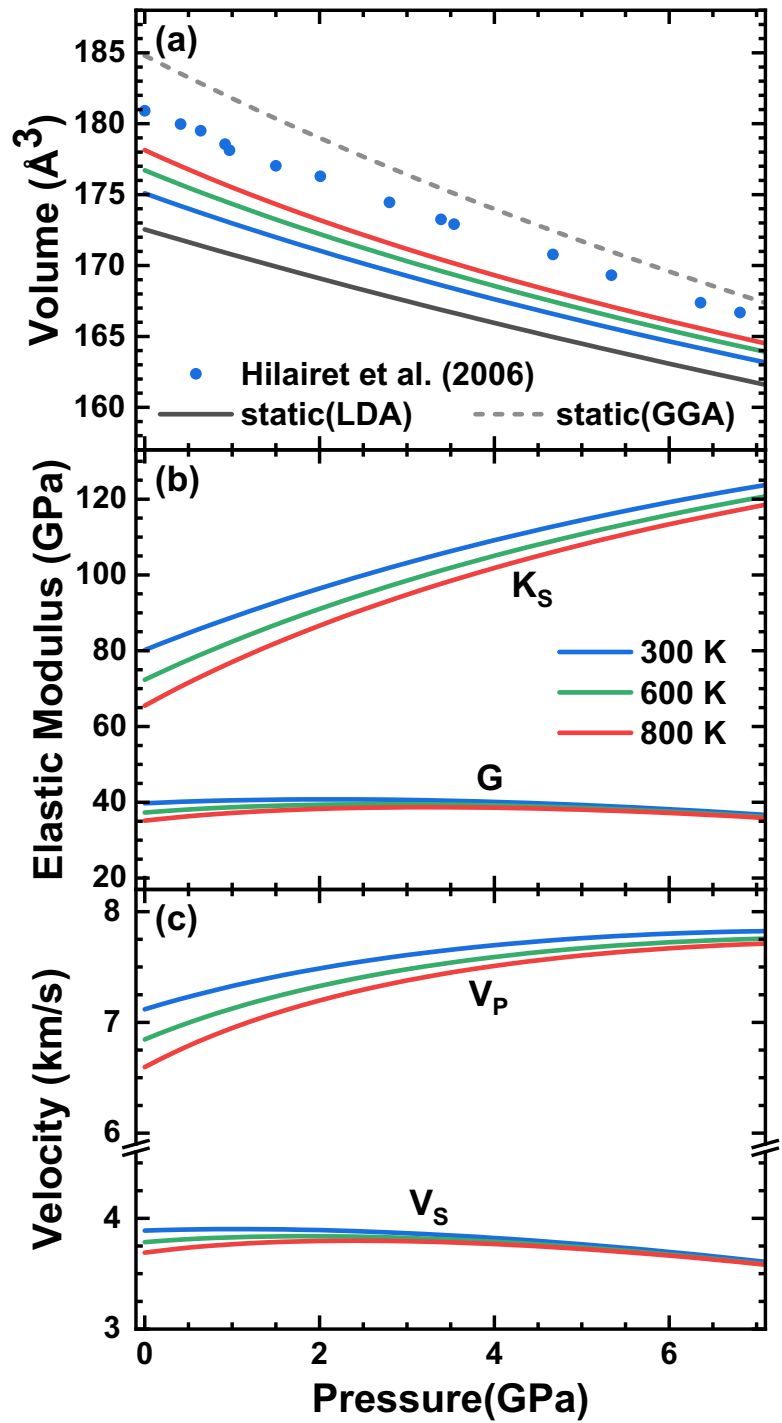

**Figure 2**. (a) Equation of states of lizardite. The solid lines represent our calculation results in LDA, the dashed line represents results within GGA, and the solid circles are experimental results in Hilairet et al. (2006), (b) adiabatic bulk modulus ($K_S$) and shear modulus ($G$), (c) compressional wave velocity ($V_P$) and shear wave velocity ($V_S$) at different temperatures and pressures.

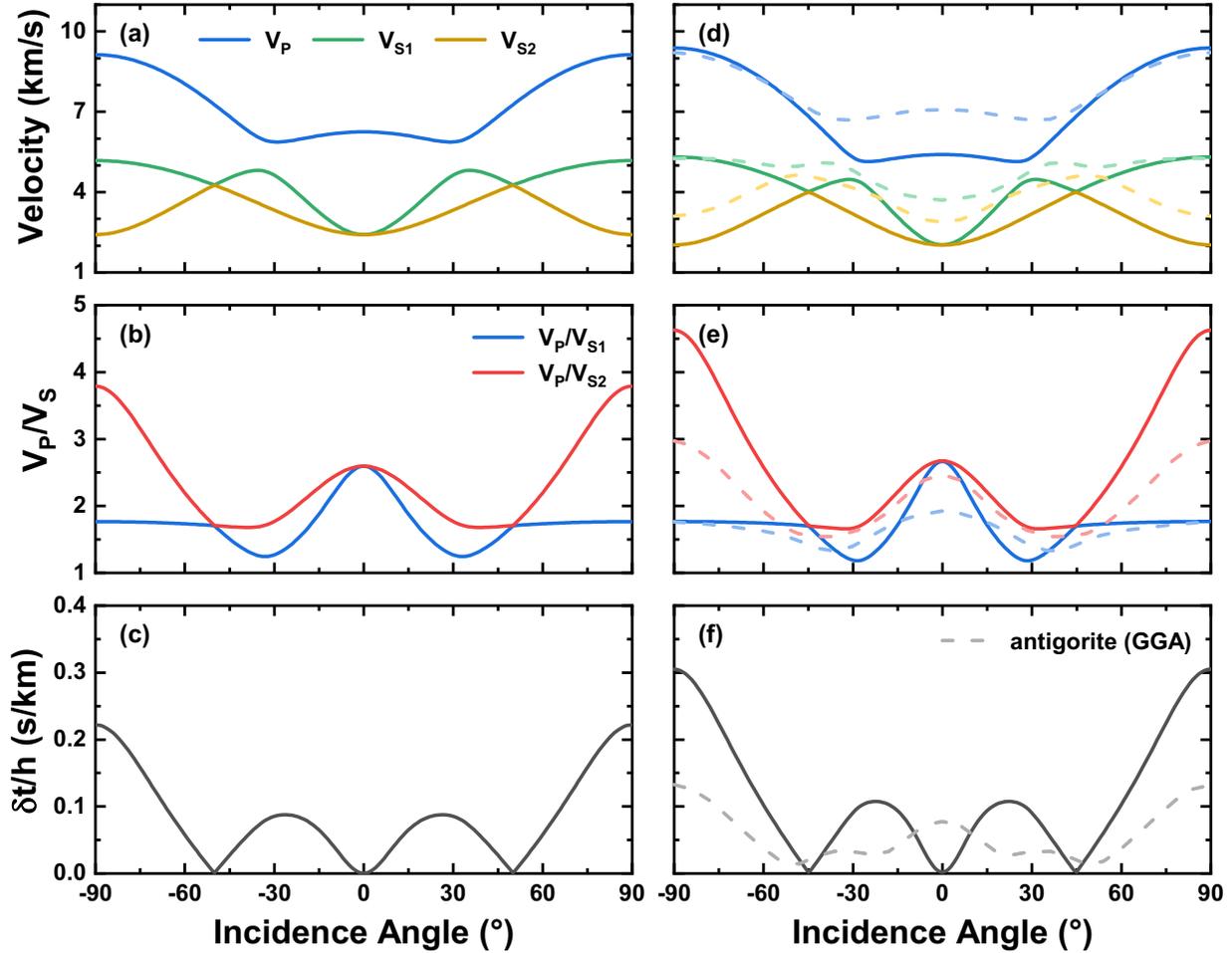

**Figure 3**. Anisotropy of lizardite: variation of (a,d) $V_P$, $V_{S1}$, and $V_{S2}$; (b,e) $V_P/V_{S1}$ and $V_P/V_{S2}$; (c,f) delay time/layer thickness ($\delta V_S/(V_{S1} \times V_{S2})$) as a function of the incidence angle between the seismic ray path and (001) plane of lizardite, at 300 °C and 0.6 GPa, corresponding to 20 km depth at slab conditions. (a-c) results within LDA; (d-f) results within GGA. The anisotropy of antigorite within GGA (Mookherjee & Capitani, 2011) is also plotted in dashed lines for comparison.

**Table 1.** Volume, elastic constants and velocities of lizardite at 0 GPa

| | $V_0(\text{Å}^3)$ | $C_{11}(\text{GPa})$ | $C_{33}(\text{GPa})$ | $C_{12}(\text{GPa})$ | $C_{13}(\text{GPa})$ | $C_{14}(\text{GPa})$ | $C_{44}(\text{GPa})$ | $C_{66}(\text{GPa})$ | $V_P(\text{km/s})$ | $V_S(\text{km/s})$ |
|---|---|---|---|---|---|---|---|---|---|---|
| *Static (LDA)* | 172.54 | 227.33 | 124.74 | 82.28 | 27.27 | 0.38 | 16.24 | 73.02 | 7.31 | 3.89 |
| *Static (GGA)* | 184.81 | 215.72 | 59.68 | 74.22 | 8.11 | 1.54 | 10.63 | 70.50 | 6.46 | 3.63 |
| *300 K* | 175.09 | 225.19 | 104.27 | 79.62 | 20.66 | 0.16 | 16.03 | 73.02 | 7.12 | 3.89 |
| LDA[a] | 172.2 | 222.5 | 104.6 | 75.2 | 18.9 | 2.2 | 17.1 | 73.7 | 7.05 | 3.90 |
| LDA[b] | 170.76 | 235.61 | 118.16 | 85.96 | 25.05 | 2.69 | 20.92 | 74.83 | 7.42 | 4.09 |
| GGA[b] | 186.34 | | | | | | | | | |
| GGA[c] | | 245 | 23 | 50 | 31 | 0 | 11.6 | 97.5 | 6.20 | 3.66 |
| GGA[d] | 182.3 | 212.6 | 57.3 | 73.3 | 8.5 | 1.3 | 11.6 | 69.7 | 6.40 | 3.63 |
| *GULP[e]* | 184.0 | 229.08 | 45.838 | 89.044 | 13.558 | 4.6025 | 12.765 | 70.017 | 6.49 | 3.65 |
| *Expt.[f]* | 180.92 | | | | | | | | | |

Predicted volume, elastic constants and velocities of lizardite at 300 K in LDA and at static conditions within both LDA and GGA, compared with previous experimental studies and static calculations at 0 GPa. Data sources: a: Ghaderi et al. (2015). b: Mookherjee and Stixrude (2009). c: Reynard et al. (2007). d: Tsuchiya et al. (2013). e: Auzende et al. (2006). f: Hilairet et al. (2006).